\documentclass{aa}

\usepackage{graphicx}
\usepackage{txfonts}
\usepackage{epsfig,latexsym,longtable,lscape}

\usepackage{natbib}
\bibpunct{(}{)}{;}{a}{}{,}

\def\xmm{XMM-Newton}
\newcommand{\nh}{N$_{\rm H}$}
\newcommand{\lx}{\hbox{L$_{\rm x}$}}

\newcommand{\fx}{\hbox{F$_{\rm x}$}}
\newcommand{\expo}[1]{$\times 10^{#1}$}
\newcommand{\oexpo}[1]{$10^{#1}$}

\newcommand{\hcm}[1]{$\times 10^{#1}$ cm$^{-2}$}

\def\SNR{\mbox{{SNR~J0528--6714}}}
\def\HPnum{\mbox{{[HP99]\,498}}}

\newcommand{\SII}{[S\,{\sc ii}]}
\newcommand{\OIII}{[O\,{\sc iii}]}
\newcommand{\Halpha}{H${\alpha}$}
\newcommand{\D}{$^\circ$}
\def\p0{\phantom{0}}

\def\it{\sl}

\begin{document}

   \title{Multi-frequency study of Local Group Supernova Remnants}
   \subtitle{The curious case of the Large Magellanic Cloud SNR~J0528--6714 }
   \author{E.J.~Crawford\inst{1} \and  M.D.~Filipovi\'c\inst{1,2} \and F.~Haberl\inst{2} \and W.~Pietsch\inst{2} \and J.L.~Payne\inst{1} \and A.Y.~De~Horta\inst{1}}

    \institute{University of Western Sydney, Locked Bag 1797, Penrith South DC,
    NSW 1797, Australia\\
        e-mail:  e.crawford@uws.edu.au, m.filipovic@uws.edu.au,
        a.dehorta@uws.edu.au
        \and
        Max-Planck-Institut f\"ur extraterrestrische
    Physik, Giessenbachstrasse, D--85741 Garching, Germany\\
    e-mail: fwh@mpe.mpg.de, wnp@mpe.mpg.de
       }
% \abstract{}{}{}{}{}
% 5 {} token are mandatory

  \abstract
    {The SNRs known in the Local Group show a variety of morphological structures that are
    relatively uncorrelated in the different wavelength bands. This variety is probably caused by the 
    different conditions in the surrounding medium with which the remnant interacts.}
    {Recent ATCA, XMM-Newton and MCELS observations of the Magellanic Clouds (MCs) cover a number of new and known SNRs which
    are poorly studied,  such as \SNR. This particular SNR exhibits luminous radio-continuum emission, but is one of the unusual and rare cases without detectable optical and very faint X-ray emission (initially detected by ROSAT and listed as object \HPnum). 
    We used new multi-frequency radio-continuum surveys and new optical
    observations at \Halpha, \SII\ and \OIII\ wavelengths, in combination with XMM-Newton X-ray data, to investigate
    the SNR properties and to search for a physical explanation for the unusual appearance of this SNR.}
    {We analysed the X-ray and Radio-Continuum spectra and present multi-wavelength morphological studies of this SNR.}
    {We present the results of new moderate resolution ATCA observations of \SNR. We found that this object is a
    typical older SNR with a radio spectral index of $\alpha$=--0.36$\pm$0.09 and a diameter of D=52.4$\pm$1.0~pc. 
    Regions of moderate and somewhat irregular polarisation were detected which are also indicative of an older 
    SNR. Using a non-equilibrium ionisation collisional plasma model to describe the X-ray spectrum, we find 
    temperatures kT of 0.26~keV for the remnant. The low temperature, low surface brightness, and large 
    extent of the remnant all indicate a relatively advanced age. The near circular morphology indicates a Type Ia 		event.}
    {Our study revealed one of the most unusual cases of SNRs in the Local Group of galaxies -- 
    a luminous radio SNR without optical counterpart and, at the same time, very faint X-ray 
    emission. While it is not unusual to not detect an SNR in the optical, the combination of faint 
    X-ray and no optical detection makes this SNR very unique.}

%
%\titlerunning{New XMM-Newton SNRs in the SMC}
%\authorrunning{M.D. Filipovi\'c et al.}
\maketitle
%
%________________________________________________________________
\section{Introduction}

The Large Magellanic Cloud (LMC), with its low foreground absorption and relative proximity of $\sim$50~kpc  \citep{2008MNRAS.390.1762D}, offers the ideal laboratory for the detailed study of a complete sample of objects such as supernova remnants (SNRs). The proximity enables detailed spatial studies of the remnants, and the accurately known distance allows for analysis of the energetics of each remnant. In addition, the wealth of wide-field multiwavelength data available, from radio maps to optical emission-line images and broad-band photometry to global X-ray mosaics, provides information about the contexts and environments in which these remnants are born and evolve.

It is possible to obtain a relatively complete sample of SNRs in the LMC and not only study the global properties of the sample but also study the subclasses in detail (e.g., sorted by \mbox{X-ray} and radio morphology, diameter or by type of the supernova progenitor). Toward this goal, we have been studying SNRs in the Magellanic Clouds (MCs) in greater detail using combined optical, radio, and X-ray observations. Today we know over 40 confirmed SNRs in the LMC and another 35-40 candidates \citep{2008MNRAS.383.1175P}. 

Here, we report on multi-frequency observations of a previously known and intriguing LMC supernova remnant. \SNR\ was initially suggested as a candidate by \citet{1984PASAu...5..537T}. \citet{1985ApJS...58..197M} classified it as an SNR based on observations made with the Molonglo Radio Observatory, at 36\,cm ($\nu$=843~MHz), and they noted no optical counterpart. \citet{1998A&AS..130..421F} added further confirmation, with a set of radio-continuum observations (with the Parkes telescope) over a wide frequency range and estimated a steep spectrum with powerlaw index \mbox{$\alpha=-0.79\pm0.15$.} \citet{2006ApJS..165..480B} observed \SNR\ but reported no detection at far-ultraviolet (FUV) wavelengths. \citet{1999A&AS..139..277H} associated an X-ray counterpart to this radio SNR from {\it ROSAT} PSPC observations. According to the entry number in their catalogue, the ROSAT source is named \HPnum.

\section{Observational Data}
 \label{observations}

\subsection{The ATCA radio-continuum observations}

We observed \SNR\ with the Australia Telescope Compact Array (ATCA) on 6$^\mathrm{th}$ April 1997, with an array configuration EW375, at wavelengths of 6 and 3~cm ($\nu$=4790 and 8640~MHz). The observations were done in so called ``snap-shot'' mode, totaling $\sim$1 hour of integration over a 12 hour period. Source 1934-638 was used for primary calibration and source 0530-727 was used for secondary calibration. The \textsc{miriad} \citep{2006Miriad} and \textsc{karma} \citep{2006Karma} software packages were used for data reduction and analysis. More information about the observing procedure and other sources observed during this session can be found in \citet{2007MNRAS.378.1237B}, \citet{2008SerAJ.177...61C,2008SerAJ.176...59C} and \citet{2009SerAJ.179...55C}.

Baselines formed with the $6^\mathrm{th}$ ATCA antenna were excluded, as the other five antennas were arranged in a compact configuration. The 6\,cm image (Fig.~\ref{fig-6cm}) has a resolution of 39\arcsec$\times$31\arcsec\ at position angle 50\D\ and the r.m.s noise is estimated to be 0.4~mJy/beam. Due to the signal to noise restrictions and the size of the remnant, no reliable image could be prepared at 3\,cm. 

We also used all available radio-continuum images of the LMC. These are composed of observations at several radio frequencies having moderate resolution at 36\,cm \citep[$\nu$=843~MHz, MOST;][]{1991IAUS..148..114T} and 20\,cm \citep[$\nu$=1400~MHz, ATCA;][]{ 2007MNRAS.382..543H}.

\subsection{\xmm\ observations and data reduction}

\xmm\ serendipitously observed \SNR\ on 28$^\mathrm{th}$ January 2007, for a total of about 20 ks (observation ID 0402000601) at an off-axis angle of $\sim$6\arcmin. The observation was performed with the EPIC instruments \citep[PN and two MOS cameras, ][]{2001A&A...365L..18S,2001A&A...365L..27T} in imaging read out mode. Thin optical blocking filters were used to optimise observations of the target, the supersoft X-ray source candidate RX\,J0529.4$-$6713 \citep{2008A&A...482..237K} which is associated with the planetary nebula SMP\,L69. The data were analysed using the analysis package XMMSAS version 8.0.0. After removal of intervals with high background activity, we obtained a net exposure time of 17.3 ks (EPIC-PN).

For morphology studies in comparison with radio-continuum and optical images \xmm\ EPIC images were produced in the standard energy bands 0.2$-$0.5~keV, 0.5$-$1.0~keV, 1.0$-$2.0~keV and 2.0$-$4.5~keV. The images show faint diffuse emission at the position of the SNR and a point-like source near its centre at RA(J2000)=5$^h$28$^m$17.9$^s$, DEC(J2000)=--67\degr14\arcmin 01\arcsec\ (statistical error 1\arcsec, systematic uncertainty 2--3\arcsec). 

To investigate if the point-like source is related to the SNR we extracted spectra for both sources separately. For the point-like source we used a circular extraction region with 20\arcsec\ radius. For the SNR the outer radius of the extraction region (centre at RA(J2000)=5$^h$28$^m$19.8$^s$, DEC(J2000)=--67\degr14\arcmin 21\arcsec) was 100\arcsec, excluding a circle around the point-like source with radius 25\arcsec. EPIC X-ray spectra were extracted for PN (single + double pixel events, corresponding to a PATTERN 0$-$4 selection) and MOS (PATTERN 0$-$12), excluding bad CCD pixels and columns (FLAG~0). We used {\sc XSPEC}\footnote{{http://heasarc.gsfc.nasa.gov/docs/xanadu/xspec/}} version 12.5.1 for spectral modelling. {The three EPIC spectra from the SNR} were fitted simultaneously, allowing only a renormalisation factor to account for cross-calibration uncertainties between the instruments and area losses for the MOS extraction as the region covered by the SNR is near the edge of the inner MOS CCD in both cameras. To account for photo-electric absorption by interstellar gas, two hydrogen column densities were used. The first represents the foreground absorption in the Milky Way, fixed at 6\hcm{20} assuming elemental abundances of \citet{2000ApJ...542..914W}. The second considers the absorption in the LMC \citep[with metal abundances set to 0.5 solar as typical of the LMC; ][]{1992ApJ...384..508R}. The statistical quality of the spectra was sufficient to fit one-component thermal plasma emission models. For the supernova remnant we used a single-temperature non-equilibrium ionisation collisional plasma (NEI) model \citep[in XSPEC, see][and references therein]{2001ApJ...548..820B} with metal abundances fixed to 0.5~solar, yielding acceptable $\chi^2$ values. Results from similar modelling of X-ray spectra from SNRs in the SMC were published by \citet{2004A&A...421.1031V} and \citet{2008A&A...485...63F}, allowing direct comparison.

\subsection{The MCELS optical surveys of the LMC}

The Magellanic Cloud Emission Line Survey (MCELS) was carried out from the 0.6~m University of Michigan/CTIO Curtis Schmidt telescope, equipped with a SITE $2048 \times 2048$\ CCD, which gave a field of 1.35\degr\ at a scale of 2.4\arcsec\,pixel$^{-1}$. Both the LMC and SMC were mapped in narrow bands corresponding to \Halpha, \OIII\ ($\lambda$=5007\,\AA), and \SII\ ($\lambda$=6716,\,6731\,\AA), plus matched red and green continuum bands that are used primarily to subtract most of the stars from the images to reveal the full extent of the faint diffuse emission. All the data have been flux-calibrated and assembled into mosaic images, a small section of which is shown in Fig.~\ref{fig-opt}. Further details regarding the MCELS are given by \citet{2006NOAONL.85..6S}, Winkler et al. (in preparation) and at http://www.ctio.noao.edu/mcels/.

Additional observations were attempted on 25$^\mathrm{th}$ September 2009,  using the 1.9-meter telescope and Cassegrain spectrograph at the South African Astronomical Observatory (SAAO) in Sutherland. While the observing conditions were good, we did not detect any emission from \SNR{}.

\section{Results}
 \label{results}

\subsection{Radio-continuum}
\begin{figure*}
  \begin{center}
  \includegraphics[angle=-90,width=170mm,clip=]{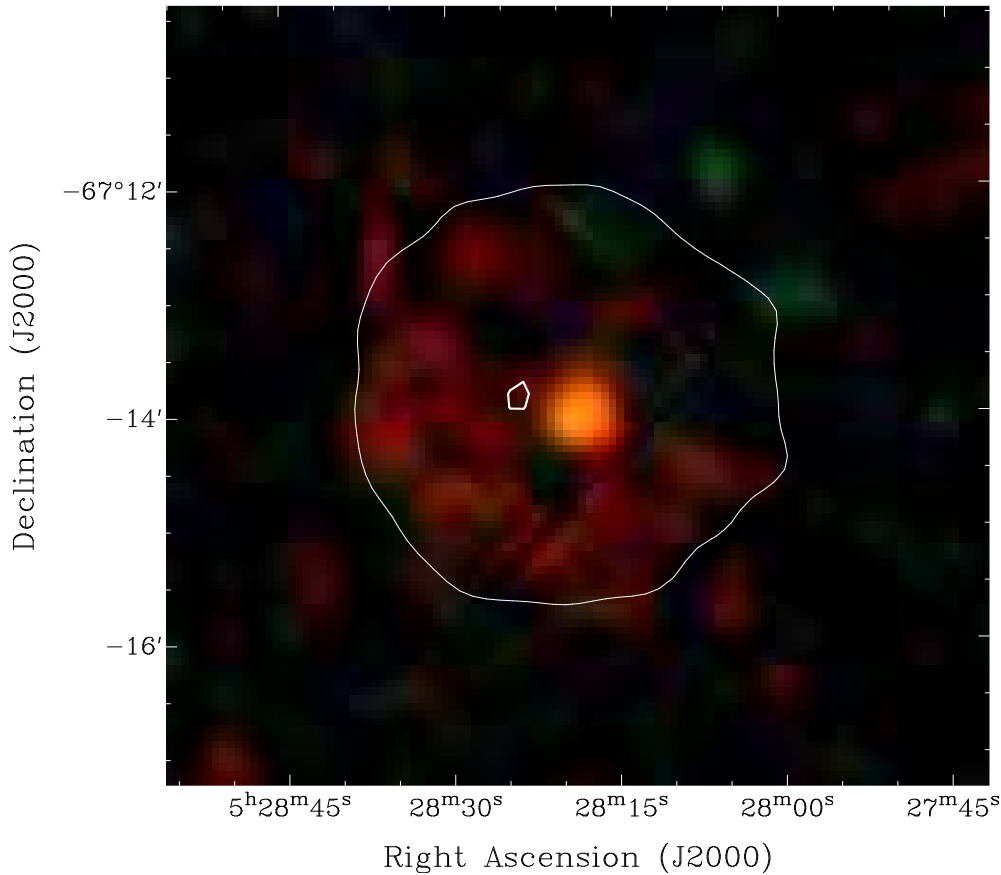}
  \end{center}
  \caption{\xmm\ false colour image (red corresponds to X-ray intensity in the 0.2$-$1.0~keV band, green to 1.0$-$2.0~keV and blue to 2.0$-$4.5~keV) of \SNR\ overlaid with an ATCA 6\,cm 3$\sigma$ (1.2~mJy) and 10$\sigma$ (4.0~mJy) radio-continuum contours.} 
  \label{fig-xray-radio}
\end{figure*}

The remnant has a circular appearance centered at RA(J2000)=5$^h$28$^m$18.5$^s$, DEC(J2000)=--67\degr13\arcmin 49.2\arcsec\ with a diameter of 216\arcsec$\pm$5\arcsec\ (52.4$\pm$1.0~pc). This is reasonably consistent with its X-ray extent as indicated in Fig.~\ref{fig-xray-radio} where the 3$\sigma$ radio-continuum contour is drawn on an X-ray image. The point-like X-ray source is located very close to the centre of the radio emission while the {diffuse} X-ray emission is brighter on the south-eastern side.

\begin{table}
\begin{center}
\caption{Integrated Flux Density of \SNR.}
\label{tab-flux}
\begin{tabular}{cccccccc}
\hline\noalign{\smallskip}
\hline\noalign{\smallskip}
Frequency (MHz)      & 843         & 1400  &  4800  \\
Wavelength (cm)      & 36         & 20  &  6  \\
 \noalign{\smallskip}\hline\noalign{\smallskip}
S$_\mathrm{I}$ (mJy) &        125$\pm$10     &    122$\pm$5 &          69$\pm$5          \\
 \hline\noalign{\smallskip}
 Reference           & Mathewson       & This   & This  \\
                     & et al. 1984 & work   & work  \\
                    \hline\noalign{\smallskip}
\end{tabular}
\end{center}
\end{table}

\begin{figure*} 
 \begin{center}
 \includegraphics[angle=-90,width=166mm]{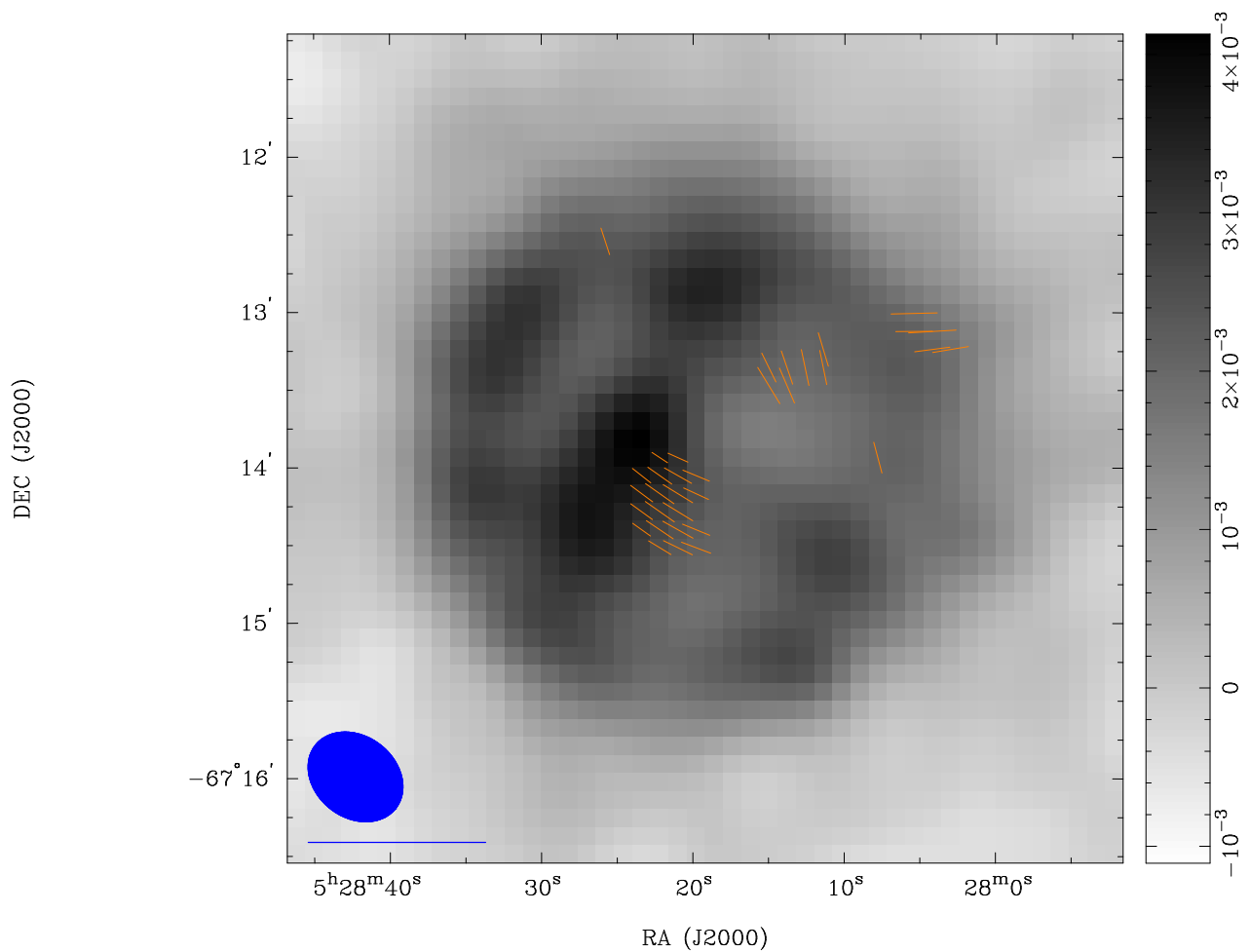}
% \vspace{9.36cm}
  \caption{ATCA observations of \SNR\ at 6~cm (4790~MHz) overlaid with fractional polarised intensity. The ellipse in the lower left corner represents the synthesised beam of 39\arcsec$\times$31\arcsec, and the line below the ellipse is a polarisation vector of 100\%. The sidebar quantifies the pixel map and its units are Jy/beam.}
  \label{fig-6cm}
 \end{center}
\end{figure*}

\begin{figure}
\begin{center}
\includegraphics[angle=-90,width=88mm]{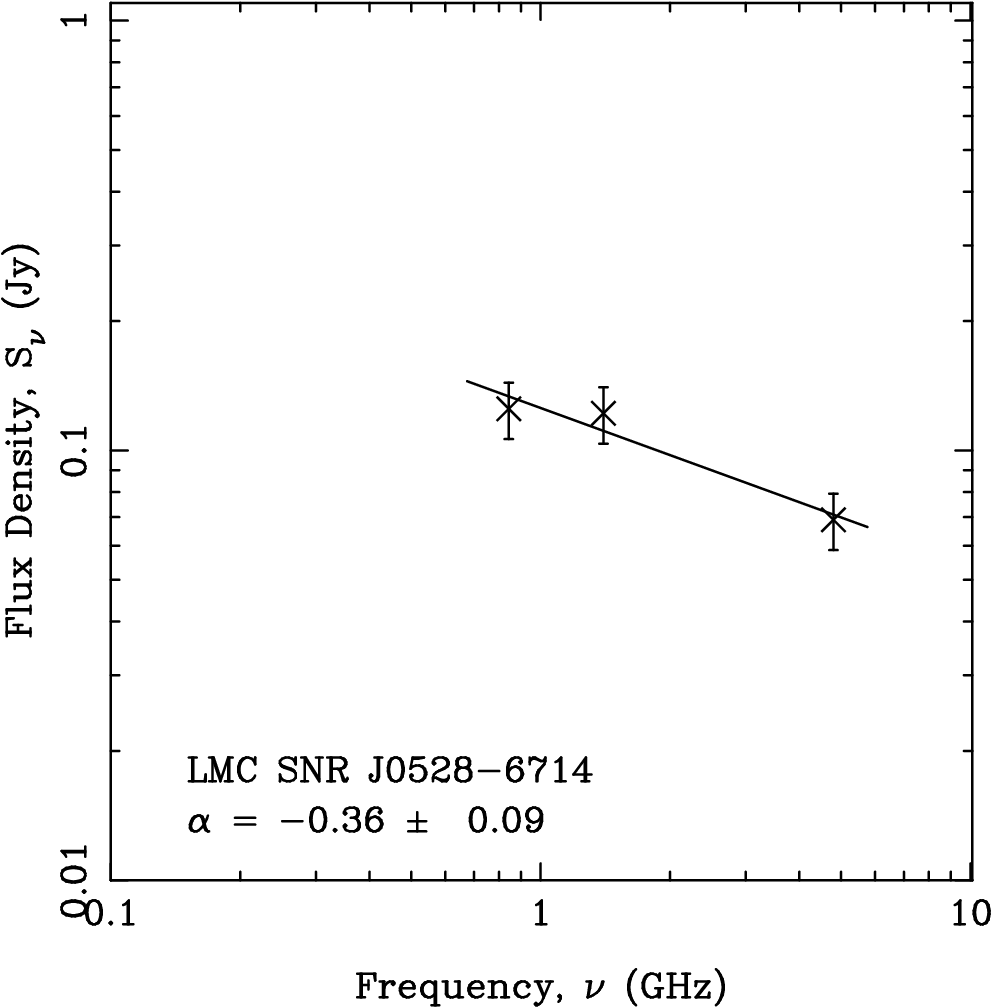}
\caption{Radio-continuum spectrum of \SNR.}
\label{fig-radspec}
\end{center}
\end{figure}

A flux density measurement was made at 6~cm, resulting in a value of $69\pm5$~mJy (Table~\ref{tab-flux}). Using the 20\,cm mosaic image described in \citet{2007MNRAS.382..543H}, we made a new estimate of the flux density of \SNR, with a value of $122\pm5$~mJy. A spectral index (defined as $S\propto\nu^\alpha$) using the flux densities in Table~\ref{tab-flux} is estimated to $\alpha$=--0.36$\pm$0.09 (Fig.~\ref{fig-radspec}), which is indicative of older SNRs \citep{1998A&AS..130..421F}. We note that the 36\,cm and 6\,cm observations are only interferometer measurements, whereas the 20\,cm observations consist of both single dish and interferometer measurements. This could lead to a flux density underestimate at these two wavelengths. Previous spectral index estimates were based solely on single dish data, \citep[\protect{$\alpha$=--0.79$\pm$0.15};][]{1998A&AS..130..421F}. We note that this is a highly confused region, and that as our new estimate uses higher resolution data than \citet{1998A&AS..130..421F}, we have been able to exclude the confusing sources from the spectral index estimate.

Linear polarization images at 6\,cm were created using stokes \textit{Q} and \textit{U} parameters. This image reveals regions of moderate and somewhat irregular polarisation within the SNR. Without reliable polarisation measurements at a second frequency we could not determine the Faraday rotation, and thus cannot deduce the magnetic field strength.

The mean fractional polarisation at 6\,cm was calculated using flux density and polarisation:
\begin{equation}
P=\frac{\sqrt{S_{Q}^{2}+S_{U}^{2}}}{S_{I}}\cdot 100\%
\end{equation}
\noindent where $S_{Q}, S_{U}$ and $S_{I}$ are integrated intensities for the \textit{Q}, \textit{U} and \textit{I} Stokes parameters. Our estimated peak value is $P \sim 20\%$ just off the centre of the remnant. Along the  shell there is a pocket of uniform polarisation, at approximately 15\% (Fig.~\ref{fig-6cm}), possibly indicating varied dynamics along the shell. This unordered polarisation is consistent with the appearance of an older SNR.

\subsection{X-Ray}

\begin{figure}
  \begin{center}
  \includegraphics[angle=-90,width=88mm,clip=]{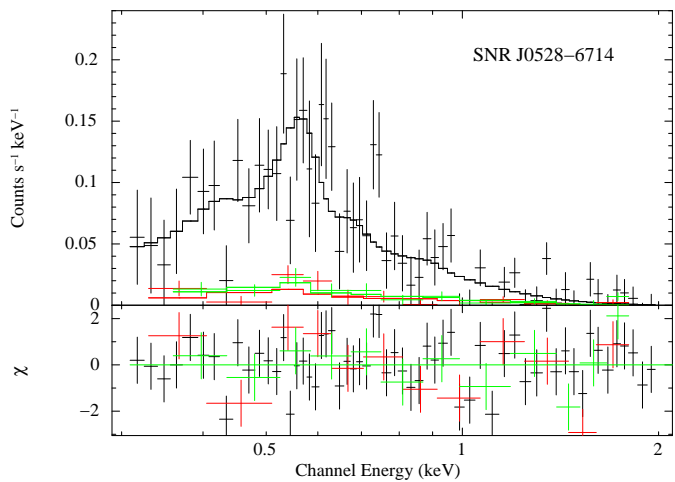}
  \end{center}
  \caption{EPIC spectra of \SNR\ (\HPnum). The best fits using a single-temperature NEI
  model are plotted as histograms (black: PN, red: MOS1, green: MOS2).}
  \label{fig-snr-spectra}
\end{figure}

The EPIC spectra of the SNR {(excluding the central source)} are plotted in Fig.~\ref{fig-snr-spectra} together with the best-fit NEI model. The derived model parameters (LMC absorption, temperature kT and ionisation time scale $\tau$) are summarised in Table~\ref{tab-fits} together with inferred fluxes and luminosities. Flux and luminosity are given for the 0.2--2~keV band, determined from the EPIC-PN spectrum (which has the best statistics with a count rate of 0.056~counts~s$^{-1}$). The intrinsic source luminosity with total \nh\ set to 0 assumes a distance of 50~kpc to the LMC. The relatively large errors in the derived LMC absorption lead to large uncertainties in the (absorption corrected) luminosities for the soft X-ray spectrum of the SNR. Therefore, we give a luminosity range derived from fits with LMC \nh\ fixed at the lower and upper 90\% confidence values. Similarly we derive a confidence range for the emission measure EM. 

Spectra extracted from the point-like source near the centre of the SNR yield low counting statistics and only a spectrum from EPIC-PN data with 5 bins between 0.3 and 2~keV could be obtained. The shape of the spectrum is similar to that of the SNR and we therefore used the same model for a fit to the PN spectrum. The resulting best fit values for absorption column density and temperature are somewhat higher than derived for the SNR, but consistent within the statistical errors. It remains unclear if the point-like source is caused by a (somewhat hotter) knot in the emission of the SNR or unrelated to it. No foreground star, which could have a similar X-ray spectrum, is found near the X-ray position.

The spectrum of \SNR\ is characterised by a high absorption and a low temperature. Considering its large extent and low surface brightness, it is most likely an older remnant. The low temperature is similar to 3 SMC SNRs \citep{2004A&A...421.1031V,2008A&A...485...63F}, and cooler than any known bright LMC SNR \citep{2004ApJ...613..948W,2005ApJ...635.1077W,2006ApJ...652.1259B,2006A&A...450..585B}.

\begin{table*}
\begin{center}
\caption{Spectral fits to the EPIC spectra of \SNR\ with an absorbed NEI model.}
\label{tab-fits}
\begin{tabular}{ccccccccc}
\hline\noalign{\smallskip}
\hline\noalign{\smallskip}
\multicolumn{1}{l}{PN exp.} &
\multicolumn{1}{c}{LMC \nh} &
\multicolumn{1}{c}{kT} &
\multicolumn{1}{c}{$\tau$} &
\multicolumn{1}{c}{\fx} &
\multicolumn{1}{c}{\lx} &
\multicolumn{1}{c}{EM} &
\multicolumn{1}{c}{$\chi^2_r$} &
\multicolumn{1}{c}{dof} \\

\multicolumn{1}{c}{[\oexpo{3} s]} &
\multicolumn{1}{c}{[\oexpo{21}cm$^{-2}$]} &
\multicolumn{1}{c}{[keV]} & \multicolumn{1}{c}{[\oexpo{8} s
cm$^{-3}$]} & \multicolumn{1}{c}{[erg cm$^{-2}$ s$^{-1}$]} &
\multicolumn{1}{c}{[erg s$^{-1}$]} & \multicolumn{1}{c}{[cm$^{-3}$]}
& \multicolumn{1}{c}{} &
\multicolumn{1}{c}{} \\

\noalign{\smallskip}\hline\noalign{\smallskip}
 17.3 & 2.9$_{-1.0}^{+2.2}$ &  0.26$_{-0.12}^{+0.17}$ &  6$_{-2}^{+3.5}$  & 1.1\expo{-13} & 9.6$_{-5.0}^{+43} $ \expo{35} & 7.4$_{-4.7}^{+65}$ \expo{58} & 1.26 & 77 \\
\noalign{\smallskip}\hline\noalign{\smallskip}
\end{tabular}
\end{center}
\end{table*}


\subsection{Optical}

\begin{figure}
  \begin{center}
  \includegraphics[width=88mm,clip=]{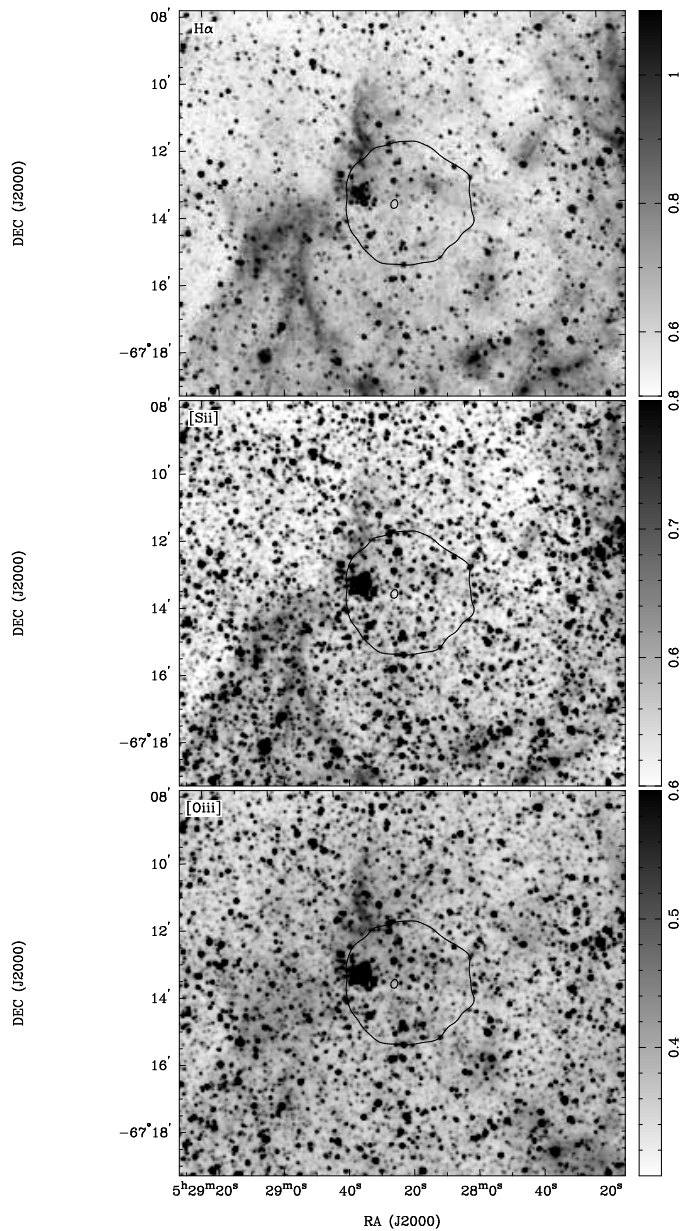}
  \end{center}
  \caption{MCELS H$\alpha$(top), \SII(Middle) and \OIII(bottom) images of \SNR\ overlaid with an ATCA 6\,cm 3$\sigma$ (1.2~mJy) and 10$\sigma$ (4.0~mJy) radio-continuum contours. }
  \label{fig-opt}
\end{figure}

There is no obvious association between any emission in the \Halpha, \SII\ or \OIII\ bands and \SNR\ (Fig.\ref{fig-opt}), confirming the earlier observation of \citet{1985ApJS...58..197M}. The high-density group of stars KMHK~943 \citep{1990A&AS...84..527K} is visible on the eastern rim of the \SNR. \citet{2006A&A...452..273M} report several B and Be stars in this region. These other objects, and their interaction with the interstellar matter (ISM), could explain the lack of an optical counterpart to \SNR. Also, we cannot exclude the possibility of a dark cloud in the line of sight.

\section{Discussion and Conclusions}
 \label{conclusions}

We have analysed our new and archival \xmm, MCELS and ATCA observations in the direction to the \SNR. We conducted the highest resolution X-ray and radio-continuum observations to date of \SNR. From radio observations we found a diameter of 52.4$\pm$1.0~pc, at the top end of the size distribution \citep{2010arXiv1003.3030B}, a radio-continuum spectral index $\alpha$=--0.36$\pm$0.09, which is indicative of an older SNR \citep{1998A&AS..130..421F}. The unordered polarisation features, most likely due to a weakening of the remnants magnetic field, also indicate an older age for \SNR. The almost perfectly circular appearance, along with the lack of radial symmetry classifies this object as a Type~Ia SN event \citep{2009ApJ...706L.106L}.

However, the most unusual feature is the lack of any optical counterpart and very faint X-ray emission \citep[and the references therein]{2008A&A...485...63F,2007AJ....134.2308B}. While it is not unusual to not detect an SNR in the optical, \citep{2001AIPC..565..267F,2004cosp...35.2434S,2005AdSpR..35.1047S}, the combination of faint X-ray and no optical detection is rare. This is in contrast to the M\,33 findings of \citet{2010arXiv1002.1839L}, where  no SNR is found with X-ray and radio emission, but not seen in optical. This may imply possible observational bias towards the optical techniques of SNR detection in external galaxies where the resolution (and therefore sensitivity) may play dominant role in SNR identification. Also, \citet{2010arXiv1002.1839L} found significant number of M\,33 SNRs with similar (or lower) X-ray energies then \SNR\ but their radio-continuum brightness is not exceedingly high as in case of \SNR.

We detect regions of moderate and somewhat irregular polarisation with maximum fractional polarisation of 15\%. Also, we argue that as with the majority of other SNRs in the MCs, this intriguing SNR is most likely in the adiabatic phase of its evolution \citep{2008MNRAS.383.1175P}.

The nature of the point-like X-ray source near the center of the remnant remains unclear. It does not coincide with the maximum radio intensity which is found somewhat off-center to the East. It is also unclear if the radio maximum is truly a point source or just enhanced emission from the shell. This, and the lack of hard X-ray emission make the explanation that \SNR\ is a Pulsar Wind Nebula (PWN) very unlikely. Also, there is no reported radio pulsar in this region \citep{2005AJ....129.1993M}.

\begin{acknowledgements}
The \xmm\ project is supported by the Bundesministerium f\"ur Wirtschaft und Technologie/Deutsches Zentrum f\"ur Luft- und Raumfahrt (BMWI/DLR, FKZ 50 OX 0001) and the Max-Planck Society. We used the Karma/MIRIAD software packages developed by the ATNF. The Australia Telescope Compact Array is part of the Australia Telescope which is funded by the Commonwealth of Australia for operation as a National Facility managed by the CSIRO. We acknowledge the many contributions of R.C. Smith, F. Winkler of CTIO and other members of the MCELS team. We were granted observation time at the South African Astronomical Observatory (SAAO) and wish to thank them for their kind help and accommodations. Travel to the SAAO was funded by Australian Government AINSTO AMNRF grant number 09/10-O-03.
{We thank the referee for their excellent comments that improved this manuscript.}
\end{acknowledgements}

\bibliographystyle{aa}
\bibliography{14767}

\end{document}